\documentclass[nocompress]{spie}  

\usepackage{amsmath,amsfonts,amssymb}
\usepackage{graphicx}
\usepackage[colorlinks=true, allcolors=blue]{hyperref}

\title{Histographs: Graphs in Histopathology}

\author[a]{Deepak Anand}
\author[a]{Shrey Gadiya}
\author[a,b]{Amit Sethi}
\affil[a]{Department of Electrical Engineering, Indian Institute of Technology Bombay, Powai, Mumbai, India}
\affil[b]{Department of Pathology, University of Illinois, Chicago, USA}

\authorinfo{Further author information: First two authors contributed equally; Send correspondence to D.A.; D.A.: E-mail: deepakanand@iitb.ac.in; S.G.: E-mail: shreygadiya@iitb.ac.in; A.S.: Email: asethi@iitb.ac.in}

\begin{document} 
\maketitle
\begin{abstract}
Spatial arrangement of cells of various types, such as tumor infiltrating lymphocytes and the advancing edge of a tumor, are important features for detecting and characterizing cancers. However, convolutional neural networks (CNNs) do not explicitly extract intricate features of the spatial arrangements of the cells from histopathology images. In this work, we propose to classify cancers using graph convolutional networks (GCNs) by modeling a tissue section as a multi-attributed spatial graph of its constituent cells. Cells are detected using their nuclei in H\&E stained tissue image, and each cell's appearance is captured as a multi-attributed high-dimensional vertex feature. The spatial relations between neighboring cells are captured as edge features based on their distances in a graph. We demonstrate the utility of this approach by obtaining classification accuracy that is competitive with CNNs, specifically, Inception-v3, on two tasks -- cancerous versus non-cancerous and in situ versus invasive -- on the BACH breast cancer dataset.
\end{abstract}

\nobreak
\section{INTRODUCTION}
\label{sec:intro}  

Research into computer-aided diagnosis in histopathology has increased significantly in the last few years with the advent of deep learning models, algorithms, and software stack that utilizes GPUs. Specifically, deep convolutional neural networks (CNNs) have been applied to histopathology images for tasks such as nucleus segmentation, gland segmentation, grading of cancer, mutation identification, and predicting recurrence of cancer. The success of CNNs for diverse tasks is based on their ability to mine task-specific features automatically from the data rather than relying on hand-crafted features. However, the starting point of the CNNs are images, a 2-D array of pixel intensities, which does not explicitly capture the inter-nuclear relationships and histopathological phenomena such as tumor infiltrating lymphocytes, glands orientations, advancing edge of the tumor at the cancer-stromal interface, etc.

Graphs that capture spatial arrangements of entities represented as vertices or sub-graphs are a natural candidate to model histopathological information extracted from a tissue image. Until now, only hand-crafted graph features have been extracted from histopathology images, such as Voronoi diagram, Delaunay triangulation, minimum spanning tree, and nuclear density calculation for solving problems such as Gleason grading in prostate cancer~\citenum{doyle2012cascaded}. Similarly, the local spatial architecture of cell clusters~\citenum{ali2013cell} and co-occurring gland angularity in localized sub-graphs~\citenum{lee2014co} have been used to predict recurrence in prostate cancer. Hand-crafted features of graphs have also been used for classifying ER+ breast cancer into high and low aggressiveness~\cite{basavanhally2013multi}. While hand-crafted features based on an expert's intuition work well for a specific dataset, their utility for a general task is a matter of investigation. The central limitation in the graph-based approach is similar to natural image domain, viz. hand-crafted features that work for one problem may not work for another. This challenge has been overcome by formulating deep neural networks to mine relevant features for a given task automatically. Similarly, automated feature mining from the graph representation of histopathological patterns should give a robust and generalized performance. 

We present a mechanism to represent histopathology images as graphs by capturing micro structure as vertex features and macro structure as edge features. This multi-scale approach bears some resemblance to how pathologists intuitively think. The proposed multi-attributed graph representation of histopathology slides is classified using a graph convolution network (GCN) that generalizes a regular CNN to operate upon graphs. We compare the classification performance of the proposed histology classification framework with the conventional patch-based CNN approach to demonstrate competitive performance on the BACH dataset for two tasks -- cancerous versus non-cancerous, and in situ vs. invasive carcinoma classification.

\section{Methods}

Our method is loosely inspired by how pathologists examine histopathology slides. Our first insight is that pathologists look at both the micro features -- such as nuclear morphology, prominence of nucleoli, margination of chromatin, and nuclear-cytoplasmic ratio -- as well as global features such as gland formation, the rupture in glands, advancing edge of the tumor at the cancer-stromal interface, etc. Thus, a valuable and interpretable representation of histopathology must capture micro features and macro spatial relationships.

Our second insight is that nuclei are not only the most prominent features at the micro-level, especially for high-powered viewing based on H\&E and several IHC stains, but they also serve as building blocks for the study of macro-level spatial structures in low-powered views by identifying their relative positioning with respect to other nuclei.

Based on the two insights above, we propose to use locations of detected nuclei as vertices of a graph of a given histopathology image. We call this graph a \emph{Histograph}. We do not distinguish between types of nuclei at the time of graph structure formation and let the distinction be learned when we compute vertex features. We capture micro-level features as vertex features based on local image descriptors, and macro-level features as edge attribute based on a mapping of Euclidean distances between neighboring nuclei. Each patient tissue image is classified by giving its Histograph as an input sample to a GCN, which is trained in a supervised manner on the training cohort of patients. GCNs allow a data-driven graph feature hierarchy to be learned without the need to specify hand-crafted features.

\begin{figure} [ht]
   \begin{center}
  \begin{tabular}{c} 
   \includegraphics[trim={0 0 0 0},clip,width=14cm]{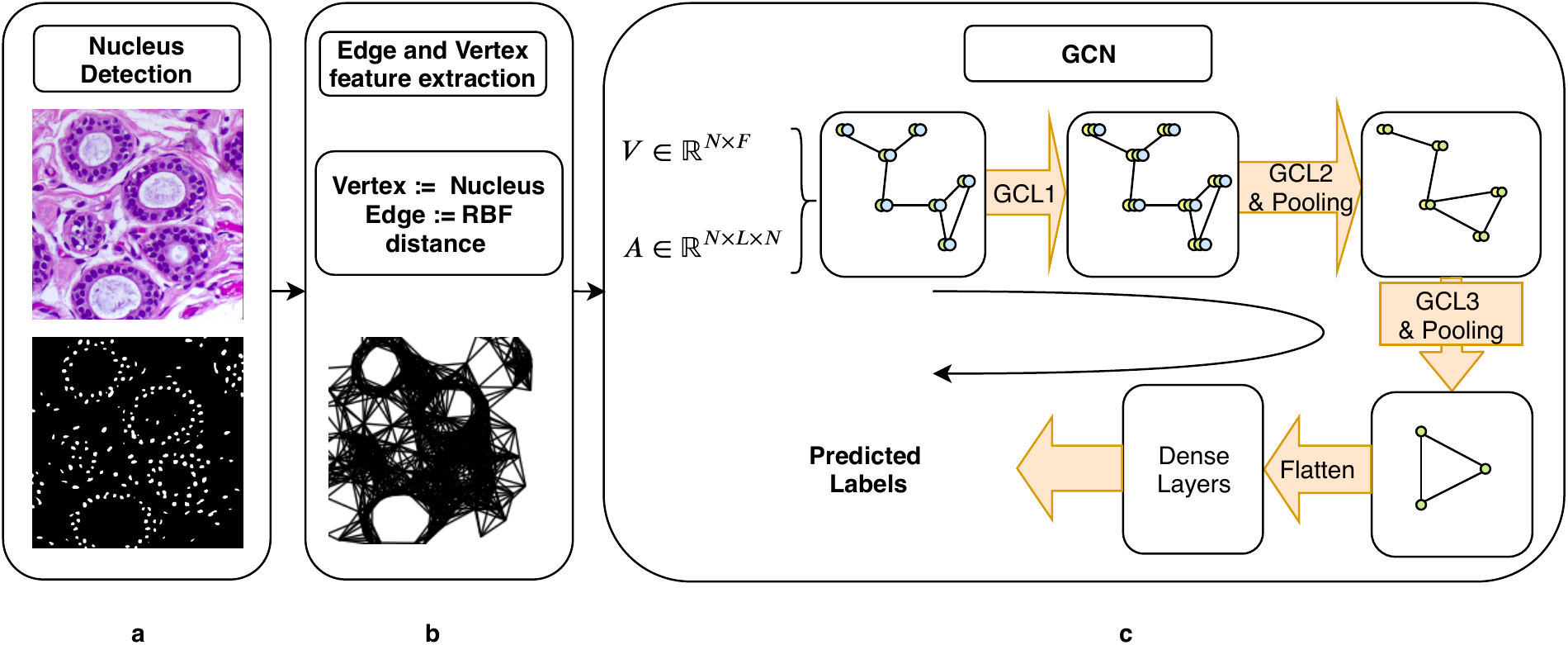}
  \end{tabular}
   \end{center}
   \caption[example] 
   { \label{fig:flow} \textbf{Flow diagram for the proposed approach:} The proposed approach consists of three stages a) nucleus detection using a fully convolutional network, b) edge and vertex features computation for the Histograph, and c) GCN-based graph classification( GCL stands for Graph convolution layers. Layers definition is based on \citenum{such2017robust} )}
   \end{figure}

The overview of the proposed approach is presented in Figure~\ref{fig:flow}. We now describe each step in more detail.

\subsection{Formation of Histograph -- a graph representation of histopathology images}
\label{subsec:graph_formation}

Given $i^{th}$ image $I_{i}$ containing $N_i$ number of detected nuclei of any kind (e.g. , epithelial, stroma, inflammatory), Histograph is a graph $\mathcal{G_i}$ using a mapping $\mathcal{F}: I_{i}\rightarrow \mathcal{G}_i(V_i,E_i)$, where the vertex set $V_i$ has cardinality $N_i$ and the edge set $E_i$ is defined over pairs of vertices from $V_i$. Micro-level features are captured as multi-attributed vertex featured extracted as local image descriptors around each nucleus. Macro-level features are functions of distances of a nucleus from other nuclei within a Euclidean distance that are stored as edge features. We describe these steps in more detail below.

\subsubsection{Nucleus detection}

To robustly detect all nuclei -- including epithelial, stroma, inflammatory -- we trained a fully convolutional neural network by augmenting a large annotated multi-organ dataset~\cite{kumar2017dataset}, such that it was robust to change in organs, magnification, disease states, and stain appearance. The architecture of the nucleus detector architecture is based on VGG-UNet, which has a VGG19 model as an encoder that is pre-trained on ImageNet, and a decoder that we trained from scratch. We separated the stains of the H\&E histopathological images, as described previously~\cite{vahadane2016}, and used the hematoxylin channel for nucleus detection. We used a switching loss that dynamically changes the relative weight between Dice, inverted Dice, and cross-entropy losses for training the network. The trained nuclei detector achieves better F1 score than the state-of-the-art models for nuclei detection.

The training dataset for the nucleus detection CNN was not used for training the GCN for tissue classification. Detected nuclei in an image of tissue to be used for training or testing tissue classification served as the vertices of the Histograph for that image.

\subsubsection{Vertex feature generation}

Once the nuclei have been detected, image features extracted from a window of size $71\times 71$ centered at the detection peak location served as multi-attributed vertex features. The vertex feature corresponding to each vertex is 438-dimensional vector comprised of the following sub-features:
\begin{itemize}
    \item \textbf{Average RGB value}: The 71$\times$71 patch is passed through a global average pooling layer to give a 3-dimensional feature vector which captures the average intensity of the staining in and around the nucleus.
    \item \textbf{Gray Level Co-occurrence Matrix (GLCM) features}: Two GLCMs of size 5$\times$5 are used, corresponding to the horizontal and vertical directions respectively. These features, when flattened, give a 50-dimensional feature corresponding to each nucleus-centric vertex capturing morphological features like clearing, alignment, and information about over-lapping or packing density of nuclei.
    \item \textbf{VGG19 features}:  The spatial extent of 71$\times$71 from first two layers of VGG19 model and then a 35$\times$35 area of next two layers are average-pooled to give a 384-dimensional feature vector.
    \item \textbf{Number of neighbors of a nuclei}: We also use the information about how connected a nucleus is in the Histograph. This information captures the density and macrostructure contribution from a particular vertex.
\end{itemize}
Concatenating these features together corresponding to each nucleus in an image we get a vertex feature matrix $V\in\mathbb{R}^{N_i\times F}$, where $N_i$ is the number of nuclei in the image and $F$ is the number of features ($F$ is 438 in our case) for each vertex. Next, we discuss the formation of edges between pairs of nuclei.

\subsubsection{Edge feature generation}

We capture the macro-structure of the tissue in the edge features of the Histograph by identifying neighbors of each nucleus that fall within a particular Euclidean distance. We used a threshold of 100 pixels based on our experience with visually being able to identify various macro-features of a tissue using just the graph structure with just enough graph density of the adjacency matrix. That is, two nuclei $n_p$ and $n_q$ are neighbors connected by an edge if $\phi(n_p,n_q)<100$, where $\phi$ is the Euclidean distance measure.

In general, one can capture multi-scale neighborhood structure by capturing multiple, say $L$, adjacency matrices based on different distance thresholds and forming an adjacency tensor $\mathcal{A}\in \mathbb{R}^{N_i \times L \times N_i}$ for the graph. For this work, we worked with $L=1$, although the GCN that we used is general enough for $L>1$ and in this case the Histograph becomes multi-attributed multi-relational graph.

\subsection{Graph convolutional network}
\label{subsec:graph_cnn}

GCN is an approach to extend the conventional convolution-based approaches to non-grid structured data. GCN defines convolution on a graph-structured dataset and facilitates geometric deep learning where the underlying structure is not necessarily based on a discrete Cartesian grid. GCNs can be categorized into two categories: a) spectral GCN that builds on graph Fourier transform and the normalized Laplacian matrix of the graph, and b) spatial GCN that builds on the formulation of spatial filter on a graph, taking $n$-hop neighbors into consideration. Due to their use of graph Fourier transform or Laplacian matrix computations, spectral approaches can only take homogeneous graph dataset as inputs where the adjacency matrix is fixed across the dataset, which would be a crippling limitation for Histographs whose number of nuclei and adjacency structure changes with the underlying tissue image. On the other hand, several spatial approaches can take heterogeneous graphs as inputs where each graph can have a different number of vertices and a different adjacency matrix. We adopted a spatial GCN called robust spatial filtering~\citenum{such2017robust} for Histograph classification.

\section{Experiments and Results}

We tested the utility of using GCNs on the Histographs using two classification tasks defined on the Breast Cancer Histology Challenge (BACH) 2018 dataset~\citenum{aresta2019bach}. This dataset consists of H\&E stained breast histology microscopy images of size   2048$\times$1536  pixels. The images are at 33X magnification as each pixel covers $0.42\mu m\times0.42\mu m$ of the tissue area. Two pathologists annotated the images, and in the case of disagreement among them, the images were discarded~\citenum{aresta2019bach}. The given images were classified into four categories: (i) normal tissue, (ii) benign lesion, (iii) in situ carcinoma, and (iv) invasive carcinoma by the two pathologists. The dataset consists of 400 images, with 100 from each category. We used a random subset of 75 images from each category for training and the remaining 25 images for testing.

The first task was to classify between non-cancerous and cancerous images, where the former class contains normal and benign tissue, while the latter class contains in situ and invasive carcinomas. Such a task corresponds to initial screening of a biopsy for malignant lesions. The second task was to classify between in situ and invasive carcinomas, which is used to decide the urgency on the extent of treatment in an oncology setting.

\subsection{Cancerous versus Non-cancerous}

We trained a GCN on Histographs of 150 non-cancerous and an equal number of cancerous images. We tested the trained GCN on Histographs of 50 images from each class. For comparison, we did the same for a popular conventional CNN architecture Inception-v3 that was pre-trained on ImageNet. The classification accuracy of both models was approximately 93\%. We want to point out that our GCN had 18 million trainable parameters, while Inception-v3 has 23 million trainable parameters.

\subsection{In situ versus Invasive}
The second task is significantly more difficult than the first, because cell-level abnormalities are similar in both in situ and invasive carcinomas, though the structure of in situ carcinoma is significantly different from that of invasive carcinoma. The former is characterized by neoplastic (tumorous) cells crowded in localized clusters that are enclosed by a distinctive thin layer of single cells, while the latter has no such defining localization of or an enclosing layer around the neoplastic cells. Thus, both the macro-structure and the micro-features are important for this task. A small patch around each nucleus cannot capture large enough context to decide the boundaries of the cancer region and the containment of neoplasia. On this task, GCN achieved an accuracy of 95\% while Inception-v3 achieved 94\%. Our GCN had 18 million trainable parameters, while Inception-v3 has 23 million trainable parameters.

\section{Conclusion and Discussion}

We proposed a framework for converting histopathological images into a graph called Histograph that is based on modeling nuclei as vertices of the graph and inter-nuclear distances as edges of the graph. This formulation enabled us to capture micro- and macro-structural properties of the histopathological slides. We also demonstrated that using GCNs on Histograph gives a competitive performance for detecting malignancy and invasiveness in BACH dataset as compared to conventional patch-based CNN. What is noteworthy is that while CNNs require aggregation of patch-level results using voting, global average pooling, or confidence thresholding schemes, a spatial GCN directly outputs a classification result on the entire graph irrespective of the size of the graph. Conversion of histology images to Histographs opens up the domain of histopathology analysis to various graph-based processing schemes, such as GCNs and geometric deep learning where the inherent structure is not necessarily Cartesian. Furthermore, we are confident that GCN visualization schemes will be able to uncover interpretable structures, such as important sub-graphs, that are responsible for tissue classification. When combined with predicting classes that are based on treatment follow-up or a genomic test, Histographs may also be used to discover new multi-scale visual biomarkers.

\acknowledgments 
 We acknowledge Nvidia for the GPU grant.  


\bibliography{report} 
\bibliographystyle{spiebib} 
\end{document}